\documentclass[aps,prl,twocolumn,showpacs,superscriptaddress,floatfix]{revtex4-1}
\usepackage{amsmath,amssymb}
\usepackage{graphicx}
\usepackage{epsfig}
\usepackage{color}
\usepackage{rotating}
\usepackage{bm}
\usepackage{setspace}

\begin{document}
\title{Detection of a superconducting phase in a two-atom layer of hexagonal Ga film\\ grown on semiconducting GaN(0001)}
\author{Hui-Min Zhang}
\affiliation{Institute of Physics, Chinese Academy of Sciences, Beijing 100190, China}
\author{Yi Sun}
\affiliation{International Center for Quantum Materials, School of Physics, Peking University, Beijing 100871, China}
\author{Wei Li}
\affiliation{State Key Laboratory of Low-Dimensional Quantum Physics, Department of Physics, Tsinghua University, Beijing 100084, China}
\author{Jun-Ping Peng}
\affiliation{Institute of Physics, Chinese Academy of Sciences, Beijing 100190, China}
\author{Can-Li Song}
\affiliation{State Key Laboratory of Low-Dimensional Quantum Physics, Department of Physics, Tsinghua University, Beijing 100084, China}
\author{Ying Xing}
\affiliation{International Center for Quantum Materials, School of Physics, Peking University, Beijing 100871, China}
\author{Qinghua Zhang}
\author{Jiaqi Guan}
\author{Zhi Li}
\affiliation{Institute of Physics, Chinese Academy of Sciences, Beijing 100190, China}
\author{Yanfei Zhao}
\affiliation{International Center for Quantum Materials, School of Physics, Peking University, Beijing 100871, China}
\author{Shuaihua Ji}
\author{Lili Wang}
\author{Ke He}
\author{Xi Chen}
\affiliation{State Key Laboratory of Low-Dimensional Quantum Physics, Department of Physics, Tsinghua University, Beijing 100084, China}
\affiliation{Collaborative Innovation Center of Quantum Matter, Beijing, China}
\author{Lin Gu}
\affiliation{Institute of Physics, Chinese Academy of Sciences, Beijing 100190, China}
\affiliation{Collaborative Innovation Center of Quantum Matter, Beijing, China}
\author{Langsheng Ling}
\author{Mingliang Tian}
\affiliation{High Magnetic Field Laboratory, Chinese Academy of Sciences, Hefei 230031, Anhui, China}
\author{Lian Li}
\affiliation{Department of Physics and Laboratory for Surface Studies, University of Wisconsin, Milwaukee, WI 53211, USA}
\author{X. C. Xie}
\affiliation{International Center for Quantum Materials, School of Physics, Peking University, Beijing 100871, China}
\affiliation{Collaborative Innovation Center of Quantum Matter, Beijing, China}
\author{Jianping Liu}
\author{Hui Yang}
\affiliation{Suzhou Institute of Nano-Tech and Nano-Bionics, Chinese Academy of Sciences, Jiangsu 215123, China}
\author{Qi-Kun Xue}
\affiliation{State Key Laboratory of Low-Dimensional Quantum Physics, Department of Physics, Tsinghua University, Beijing 100084, China}
\affiliation{Collaborative Innovation Center of Quantum Matter, Beijing, China}
\author{Jian Wang}
\email[]{jianwangphysics@pku.edu.cn}
\affiliation{International Center for Quantum Materials, School of Physics, Peking University, Beijing 100871, China}
\affiliation{Collaborative Innovation Center of Quantum Matter, Beijing, China}
\author{Xucun Ma}
\email[]{xucunma@mail.tsinghua.edu.cn}
\affiliation{Institute of Physics, Chinese Academy of Sciences, Beijing 100190, China}
\affiliation{State Key Laboratory of Low-Dimensional Quantum Physics, Department of Physics, Tsinghua University, Beijing 100084, China}
\affiliation{Collaborative Innovation Center of Quantum Matter, Beijing, China}
\date{\today}

\begin{abstract}
The recent observation of superconducting state at atomic scale has motivated the pursuit of exotic condensed phases in two-dimensional (2D) systems. Here we report on a superconducting phase in two-monolayer crystalline Ga films epitaxially grown on wide band-gap semiconductor GaN(0001). This phase exhibits a hexagonal structure and only 0.552 nm in thickness, nevertheless, brings about a superconducting transition temperature $T_{\textrm{c}}$ as high as 5.4 K, confirmed by \textit{in situ} scanning tunneling spectroscopy, and \textit{ex situ} electrical magneto-transport and magnetization measurements. The anisotropy of critical magnetic field and Berezinski-Kosterlitz-Thouless-like transition are observed, typical for the 2D superconductivity. Our results demonstrate a novel platform for exploring atomic-scale 2D superconductor, with great potential for understanding of the interface superconductivity.
\end{abstract}
\maketitle
\begin{spacing}{1.015}
Superconductivity has recently been observed in one-atomic-layer Pb \cite{Guo2004, Ozer2006, Qin2009, Zhang2010, Brun2014} and In \cite{Uchihashi2011, Yamada2013} films grown on Si(111) substrate, at the SrTiO$_3$/LaAlO$_3$ interface \cite{Reyren2007}, and in one-unit-cell thick FeSe films on SrTiO$_3$ \cite{Wang2012, Zhang2014}. This has been stimulating great attention and interest for both understanding the electron pairing in quantum confined systems and also the pursuit of emergent phases of matter in the two-dimensional (2D) systems, such as the enhancement of superconducting transition temperature $T_{\textrm{c}}$. The recent discovery of electric field induced superconductivity at SrTiO$_3$ surface \cite{Ueno2008} and in 2D MoS$_2$ crystal \cite{Ye2012} further demonstrates the feasibility of controlling 2D superconductivity via interface engineering. Thus far, however, the nature of interface or 2D superconductivity remains obscure. Preparing more hybird heterostructures with enhanced superconductivity is particularly required but experimentally challenging.

GaN, as a wide band gap and high piezo-electric constant semiconductor \cite{Bernardini1997, Ambacher1999}, is commonly used in high-speed transistors, lasers for telecommunications, and light-emitting diodes for energy efficient displays. More significantly, it has been previously shown that GaN is often wetted with one to two atomic layers of Ga atoms \cite{Xue1999, Sun2010, Northrup2000}, wherein Ga is intrinsically superconductive \cite{Gregory1966, Parr1973, Cohen1967}. Therfore, Ga/GaN may possibly serve an ideal system to search for enhanced superconductivity near their interface. In this work, by \textit{in situ }scanning tunneling microscopy/spectroscopy (STM/STS), \textit{ex situ} electrical magneto-transport and magnetization measurements, we have unambiguously demonstrated that two-monolayer (ML) Ga films (as thin as 0.552 nm) grown on GaN form a hexagonal structure and exhibit superconductivity with a $T_{\textrm{c}}$ up to 5.4 K, which differs from any previously reported stable or crystalline Ga phases \cite{Gregory1966, Parr1973, Cohen1967}. The anisotropy of critical magnetic field and Berezinski-Kosterlitz-Thouless (BKT)-like transition are observed, indicative of the 2D nature of superconductivity in 2 ML Ga/GaN(0001).

Our STM/STS experiments are conducted in a Unisoku ultrahigh vacuum low temperature STM system interconnected to a molecular beam epitaxy (MBE) chamber for film preparation. The base pressure is lower than 2 $\times$ 10$^{-10}$ Torr. All Ga films are epitaxially grown on 3 $\mu$m thick GaN(0001), which are deposited by metal organic chemical vapor deposition onto Al$_2$O$_3$(0001) substrates with a 25 nm AlN buffer layer. The substrates are cleaned by ethanol and acetone before being transferred into the MBE chamber. After degassing at $300^{\circ}$C for 3 hours, several cycles of argon ion sputtering (700 V, 2 $\times$ 10$^{-6}$ Torr) and sebsequent annealing in Ga flux are performed to remove the contaminations on the surface. Two monolayers of Ga are then epitaxially grown at $650^{\circ}$C from a high purity Ga (99.995$\%$) source with a nominal beam flux of 0.4 ML/min. Polycrystalline Pt-Ir tip is used for all STM/STS investigations. All differential conductance \textit{dI/dV} spectra are acquired using a standard lock-in technique with a bias modulation of 0.2 mV at 987.5 Hz. For \textit{ex situ} transmission electron microscope (TEM) and transport measurements, the insulating GaN substrates are used, and $\sim$ 80 nm-thick granular ($\sim$ 10 nm in size) Ag, acting as a protective and capping layer, is deposited on Ga films at 110 K before exposing the sample to the atmosphere. Note that the insulating amorphous Si capping layer has also been tried. However, we find that Si will deteriorate seriously the 2 ML Ga thin films on GaN and strongly suppress their superconductivity.

\end{spacing}
\begin{figure}[h]
\includegraphics[width=1\columnwidth]{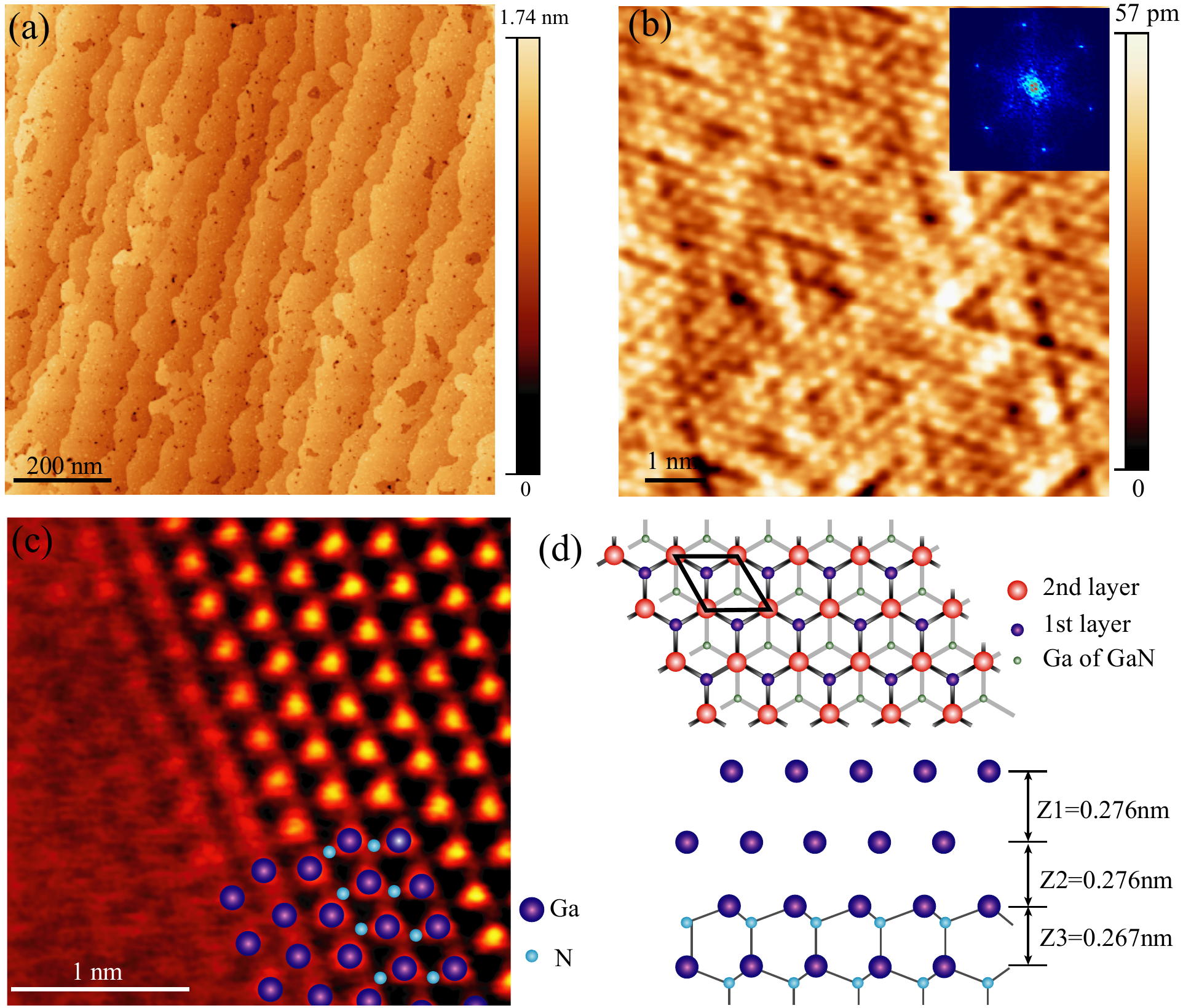}
\caption{(Color online) (a) Topographic image (3.0 V, 0.05 nA, 1 $\times$ 1 $\mu$m$^2$) of 2 ML Ga films, with a step height of 2.5 {\AA}. (b) Atomic-resolution STM image (0.22 V, 0.05 nA, 8 $\times$ 8 nm$^2$) on Ga films. The bright spots correspond to Ga atoms at the top layer, and the inset is its corresponding FFT image. (c) Cross-sectional high-angle annular dark-field image of Ag/Ga/GaN(0001) heterostructure viewed from the $[11\overline{2}0]$ crystallographic direction, showing two Ga atomic layers just above the GaN substrate. (d) Schematic top (top panel) and section (bottom panel) views of 2 ML Ga/GaN heterostructure. The average separations between various layers are Z1 = 2.76 {\AA}, Z2 = 2.76 {\AA} and Z3 = 2.67 {\AA}.
}
\end{figure}

Figure 1(a) shows the morphology of an atomically flat Ga film. The terraces, which are, on average, 150 nm wide, are separated by 2.5 {\AA} height steps, consistent with a Ga-N bilayer unit cell along the [0001] direction. Figure 1(b) depicts the atomically resolved STM and its corresponding FFT images, which exhibit a hexagonal lattice with a lattice constant of 3.18 {\AA}, close to that of the underlying GaN(0001) substrate (3.19 {\AA}). Since any previously reported stable and metastable Ga phases show either an orthorhombic or monoclinic symmetry [Table S1] \cite{Liu2004, supplementary}, the observed 2 ML Ga films with hexagonal lattice are most likely stabilized by the wurtzite structure of underlying GaN(0001) substrate, and linked to the pseudo 1 $\times$ 1 phase at room temperature. The TEM experiment reveals a sharp Ga/GaN(0001) interface and a Ga coverage of 2 ML, directly adjacent to GaN substrate [Fig.\ 1(c)]. The spacing between the two Ga layers and GaN substrate is estimated to be 0.276 nm, as schematically sketched in Fig.\ 1(d).

\begin{figure}[tb]
\includegraphics[width=1\columnwidth]{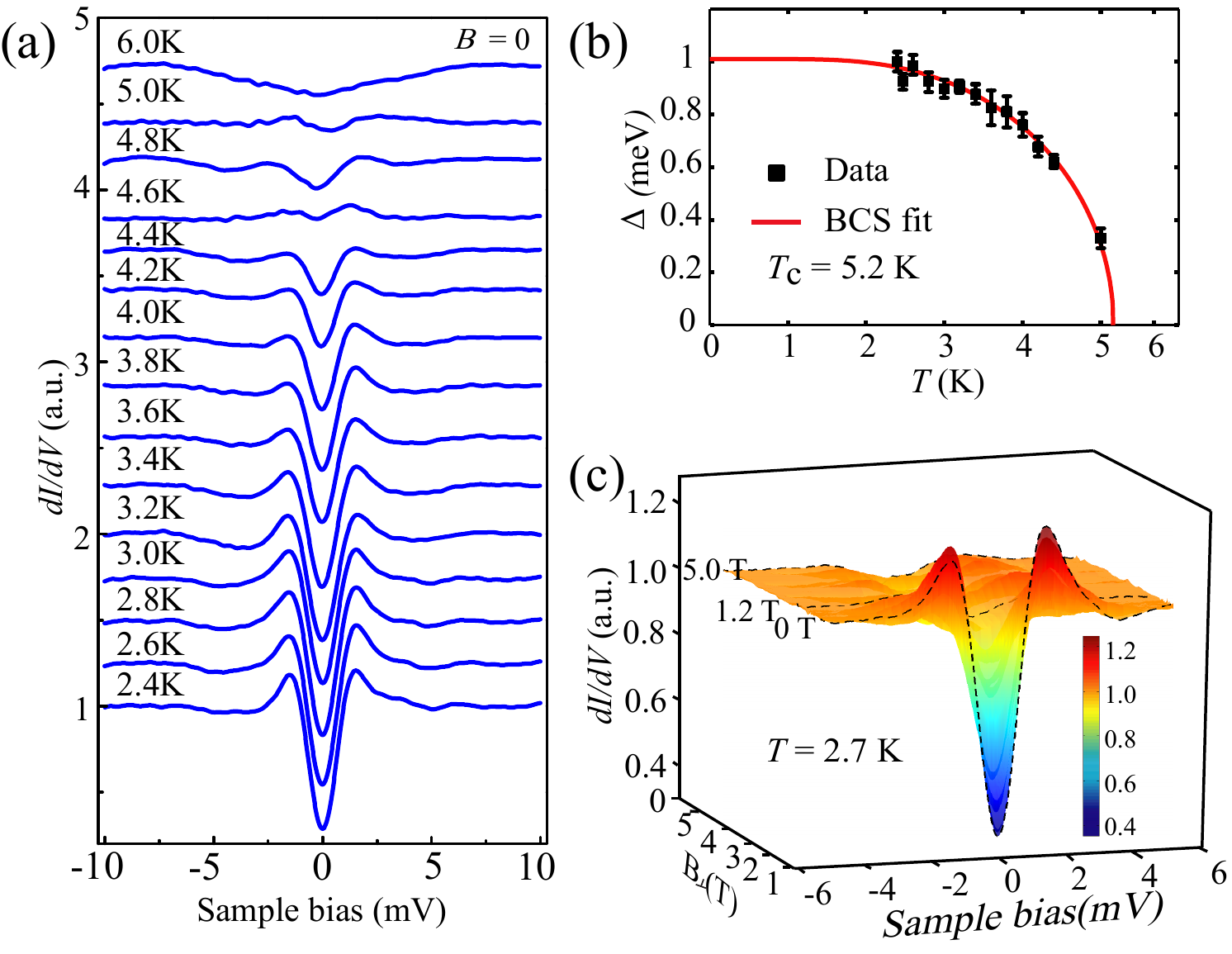}
\caption{(Color online). (a) A series of differential tunneling conductance spectra (setpoint: 10 mV, 0.1 nA) at various temperatures, normalized to the normal conductance spectrum at 10 K. (b) Temperature-dependent superconducting gap magnitude $\Delta$ (dark squares) and their best fit to BCS gap function (red curve) for 2 ML Ga films. (c) Three-dimensional plots of tunneling conductance measured at various magnetic fields at 2.7 K. Spectra measured at 0 T, 1.2 T and 5.0 T are labeled by black dashes.
}
\end{figure}

By taking differential conductance \textit{dI/dV} spectra on 2 ML Ga films at various temperatures ranging from 2.4 K to 6 K [Fig.\ 2(a)], we observe a series of temperature-dependent superconducting gaps with two clear coherence peaks at $\pm$ 1.6 meV. The measured gaps reconcile well with the well-known BCS \textit{s}-wave Dynes function with a broadening factor $\Gamma$ \cite{Dynes1978}, as illustrated in Fig.\ S1. The best fits of the data to BCS gap function \cite{Bardeen1957} yield $\Delta(0)$ = $1.01\pm0.05$ meV, $T_{\textrm{c}}\sim$ 5.2 K, and BCS ratio 2$\Delta$/\textit{k}$_{\textrm{B}}$$T_{\textrm{c}}$ =4.5 $\pm$ 0.2 ($k_{\textrm{B}}$ is the Boltzmann constant) [Figs.\ 2(b)], indicative of a strong coupling superconductor for 2 ML Ga/GaN(0001) \cite{Gregory1966}. Figure 2(c) illustrates the \textit{dI/dV} spectra as a function of the applied magnetic field normal to the sample surface ($B_\bot$). With increasing $B_\bot$, the zero bias conductance progressively increases and both the superconducting coherence peaks gradually smear out, providing the solid evidence of superconductivity in 2 ML Ga films. It is worth noting that here $T_{\textrm{c}}$ exceeds 5 K, five times higher than 1.08 K for bulk stable $\alpha$-Ga phase \cite{Parr1973, Cohen1967}.

\begin{figure}[t]
\includegraphics[width=1\columnwidth]{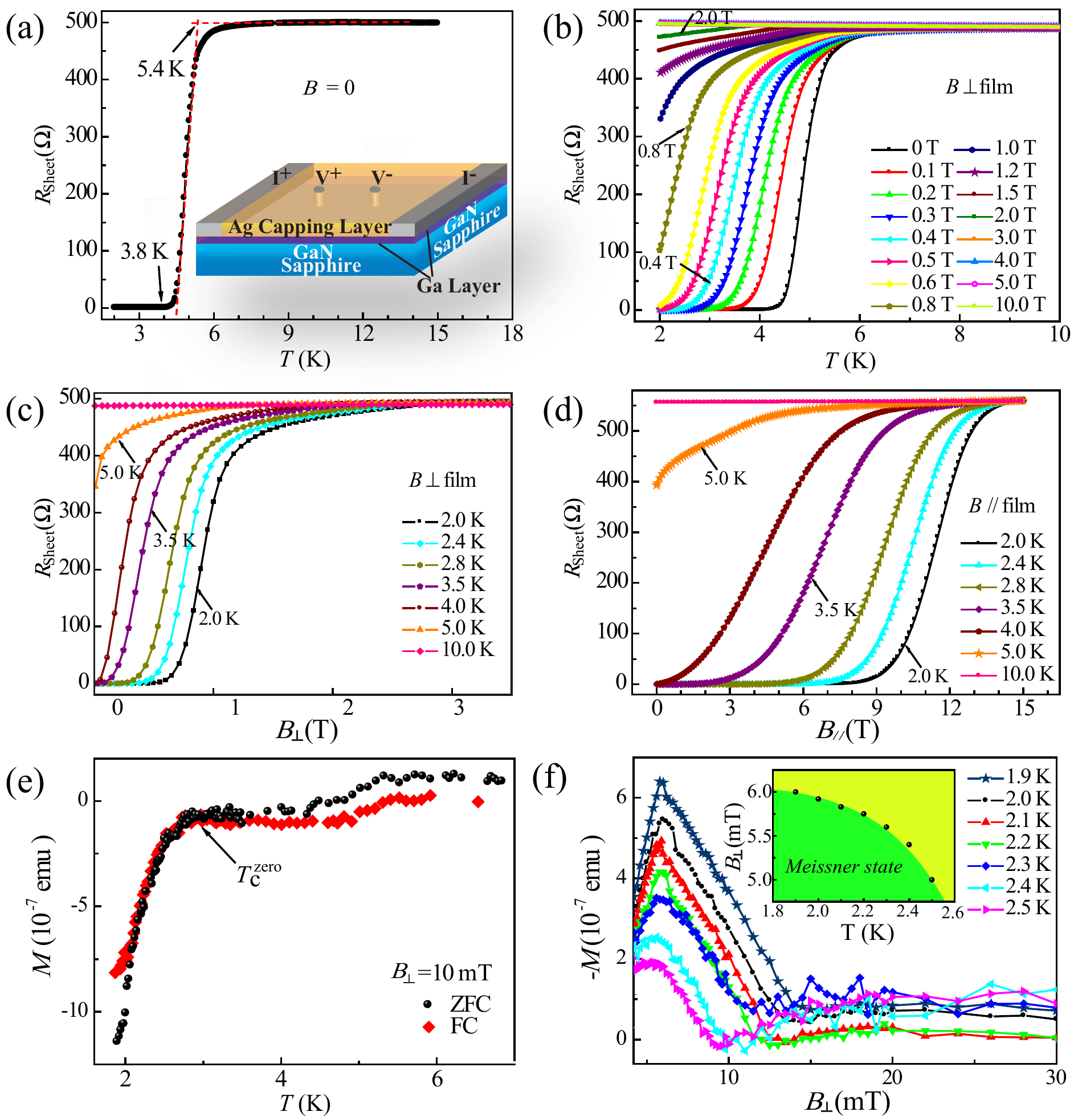}
\caption{(Color online). (a) $R_{\textrm{sheet}}$-\textit{T} curve at zero magnetic field, showing $T_{c}^{\textrm{onset}}$ = 5.4 K and $T_{c}^{\textrm{zero}}$ = 3.8 K, respectively. The inset schematically shows the diagram for all transport measurements, where indium has been used for all electrical contacts. (b) $R_{\textrm{sheet}}$-\textit{T} curves for various $B_\bot$ up to 10 T. (c, d) Magnetoresistance (c) $R_{\textrm{sheet}}$-$B_{//}$ and (d) $R_{\textrm{sheet}}$-$B_\bot$ at various temperatures ranging from 2.0 K to 10 K. (e) Temperature dependence of magnetization measured under a 10 mT  magnetic field normal to the sample surface, signaling the obvious Meissner effect. (f) Low-field \textit{M}($B_\bot$) at various temperatures from 1.9 K to 2.5 K. Note that the magnetization signal below $\sim$ 4 mT is too small to be resolved in our measurement. Inset shows the temperature dependence of $B_{\textrm{c1}}$(\textit{T}). The excitation current of 5 $\mu \textrm{A}$ is used for all $R_{\textrm{sheet}}$-\textit{T} and $R_{\textrm{sheet}}$-\textit{B} measurements throughout this paper.
}
\end{figure}

\begin{figure}[h]
\includegraphics[width=1\columnwidth]{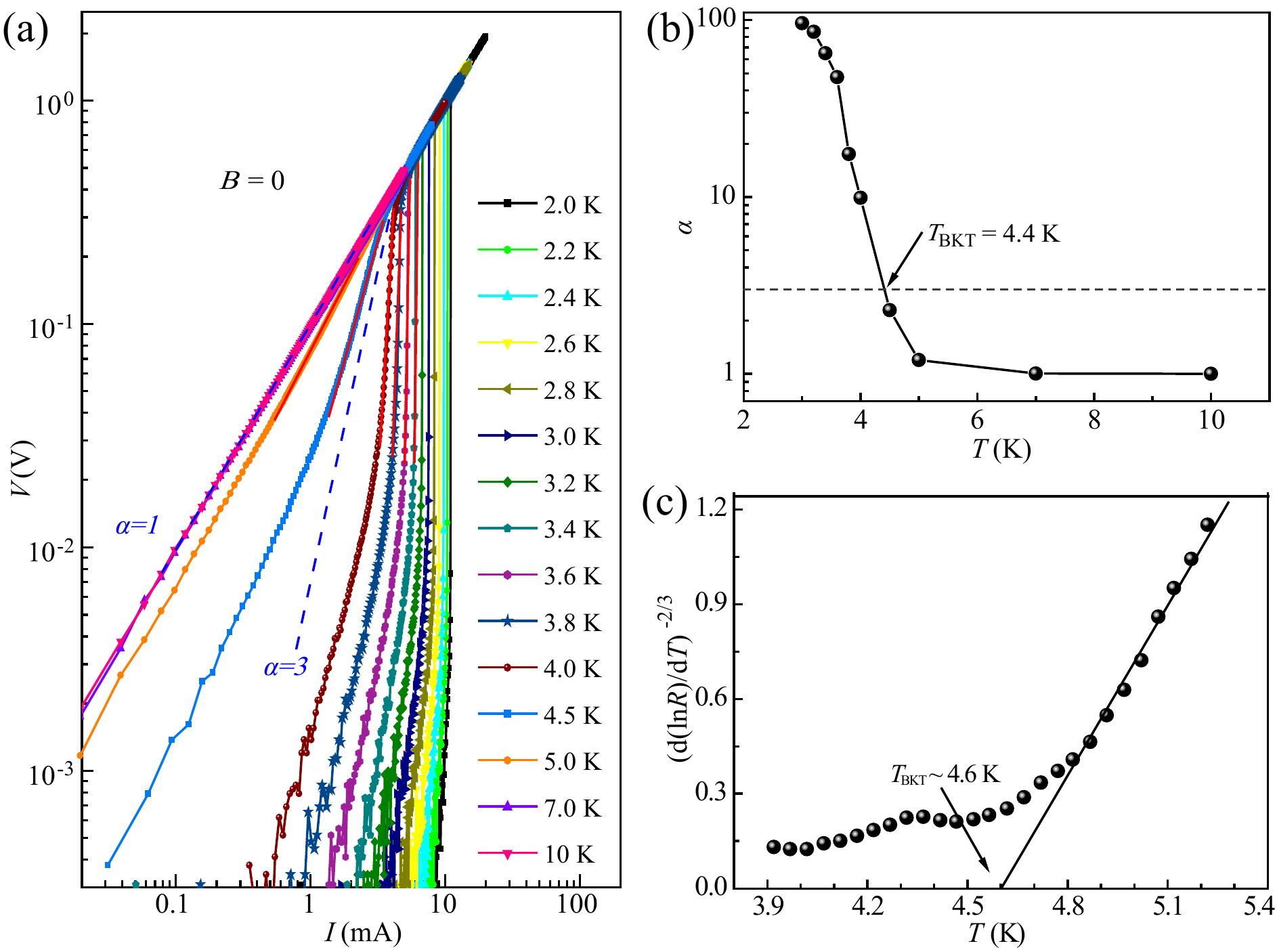}
\caption{(Color online). (a) \textit{V}(\textit{I}) characteristics at various temperatures plotted on a double-logarithmic scale at various temperatures and \textit{B} = 0 T. Two dashed blue lines indicate the \textit{V} $\sim$ \textit{I} and V $\sim I^3$ curves, respectively. (b) Plot of the exponent $\alpha$ as a function of temperature \textit{T}, extracted from the power-law fits in (a). $T_{\textrm{BKT}}$ = 4.4 K is defined as the temperature with $\alpha$ = 3. (c) $[\textrm{dln}(R_{\textrm{sheet}})/dT]^{-2/3}$ plotted as a function of temperature. The solid line depicts the expected BKT-like transition behavior with $T_{\textrm{BKT}}$ = 4.6 K.
}
\end{figure}

The high $T_{\textrm{c}}$ in 2 ML Ga/GaN(0001) has been further corroborated by our systematic transport measurements, with a schematic diagram inserted in Fig.\ 3(a). Figure 3(a) displays the sample sheet resistance ($R_{\textrm{sheet}}$) as a function of temperature at zero field, with the logarithmic-scale $R_{\textrm{sheet}}$-\textit{T} curve shown in Fig.\ S2. The superconductivity transition is immediately evident, with $T_{c}^{\textrm{onset}}$ = 5.4 K consistent with our STS measurements above. Below 3.8 K, the sample shows zero resistance within our instrumental resolution ($\pm$ 15 nV). Figure 3(b) shows the \textit{R}$_{\textrm{Sheet}}$ as a function of temperature at different $B_\bot$. The superconductivity transition gets broader and shifts to lower temperature as the field $B_\bot$ increases, as expected. In addition, magneto-transport measurements are carried out at various temperatures between 2.0 K and 10.0 K, with the fields normal [Fig.\ 3(c)] and parallel [Fig.\ 3(d)] to the sample surface, respectively. It is clearly evident that \textit{R}$_{\textrm{sheet}}$ alters differently with $B_\bot$ and $B_{//}$. For example, \textit{R}$_{\textrm{sheet}}(B_\bot)$ reaches the normal resistance at $\sim$ 3.26 T (the upper critical field $B_{\textrm{c2}}$), substantially smaller than $B_{\textrm{c}}$ = 14.8 T (the critical field) for the parallel field at 2 K. Nevertheless, both critical fields appear significantly greater than $B_{\textrm{c}}$ = 5.83 mT for bulk $\alpha$-Ga \cite{Berger2004}, which may originate primarily from the reduced dimensionality of 2 ML Ga. The small discrepancy of normal-state resistance in Figs.\ 3(c) and 3(d) stems from the sample degradation-related aging effect, because we conducted the out-of-plane field measurements firstly. The anisotropy in observed critical fields provides the first direct evidence of a typical 2D superconductor behavior for 2 ML Ga/GaN(0001). This is further solidified by analyzing the temperature dependent characteristic magnetic fields $B_\bot(T)$ and $B_{//}(T)$ \cite{Reyren2009}. Here $B_\bot(T)\propto 1-T/T_\textrm{c}$ and $B_{//}(T)\propto (1-T/T_\textrm{c})^{1/2}$ are found, highly suggestive of 2D superconductivity \cite{tinkam1996introduction} with an estimated superconducting layer thickness of $\sim$ 5.5 nm [Fig.\ S3]. Moreover, we conduct diamagnetic measurements in Fig.\ 3(e), which shows the dc magnetization as a function of temperature during the zero field cooling (ZFC) and field cooling (FC) at a perpendicular magnetic field of $B_\bot$ = 10 mT. An apparent drop appears slightly below 3.0 K, indicating the Meissner effect. The \textit{M}($B_\bot$) curves at various temperatures are shown in Fig.\ 3(f), all of which exhibit the expected linear behavior at low fields ($\leq$ 5 mT). At around $B_{\textrm{c1}}$(the lower critical field), they deviate from linearity, with temperature-dependent $B_{\textrm{c1}}$ plotted in the inset of Fig.\ 3(f). All these observations compellingly demonstrate the superconductivity in 2 ML Ga/GaN(0001).

To shed more insight into the nature of the superconductivity in 2 ML Ga/GaN(0001), Figure 4(a) shows \textit{V}(\textit{I}) characteristics at various temperatures ranging from 2 K to 10 K. A $V\sim I^{\alpha}$ power-law dependence is apparently observed near $T_{\textrm{c}}$ (red lines), with the slope equal to the exponent $\alpha$. It is found that $\alpha$ reduces with increasing temperature [Fig.\ 4(b)], consistent with a BKT-like transition \cite{Reyren2007}. The exponent $\alpha$ approaches 3 at $\sim$ 4.4 K, identified as $T_{\textrm{BKT}}$. Furthermore, close to $T_{\textrm{BKT}}$ the measured $R_{\textrm{sheet}}$ depends on temperature via $R_{\textrm{sheet}}(T)=R_{\textrm{0}}\textrm{exp}[-b(T/T_{\textrm{BKT}}-1)^{-1/2}]$, where $R_{\textrm{0}}$ and \textit{b} are material-dependent parameters \cite{Reyren2007}. This is well illustrated in Fig.\ 4(c), yielding $T_{\textrm{BKT}}$ =4.6 K, which matches with the above $\alpha$ exponent analysis. Note that the transition observed here is not sharp as theoretically expected, quite similar to the previous experimental reports in the LaAlO$_3$/SrTiO$_3$ interface superconducting system \cite{Reyren2007} and other quasi-2D superconducting systems, such as FeSe films on SrTiO$_3$ [10] and Pb films on Si(111) \cite{Zhao2013}. The broad transition might result from the finite size effect or interface effect, which has been demonstrated to play an important role in the non-freestanding quasi-2D superconducting systems \cite{Guo2004, Ozer2006, Qin2009, Zhang2010, Brun2014, Uchihashi2011, Yamada2013, Reyren2007, Wang2012, Zhang2014}.

\begin{figure}[h]
\includegraphics[width=0.55\columnwidth]{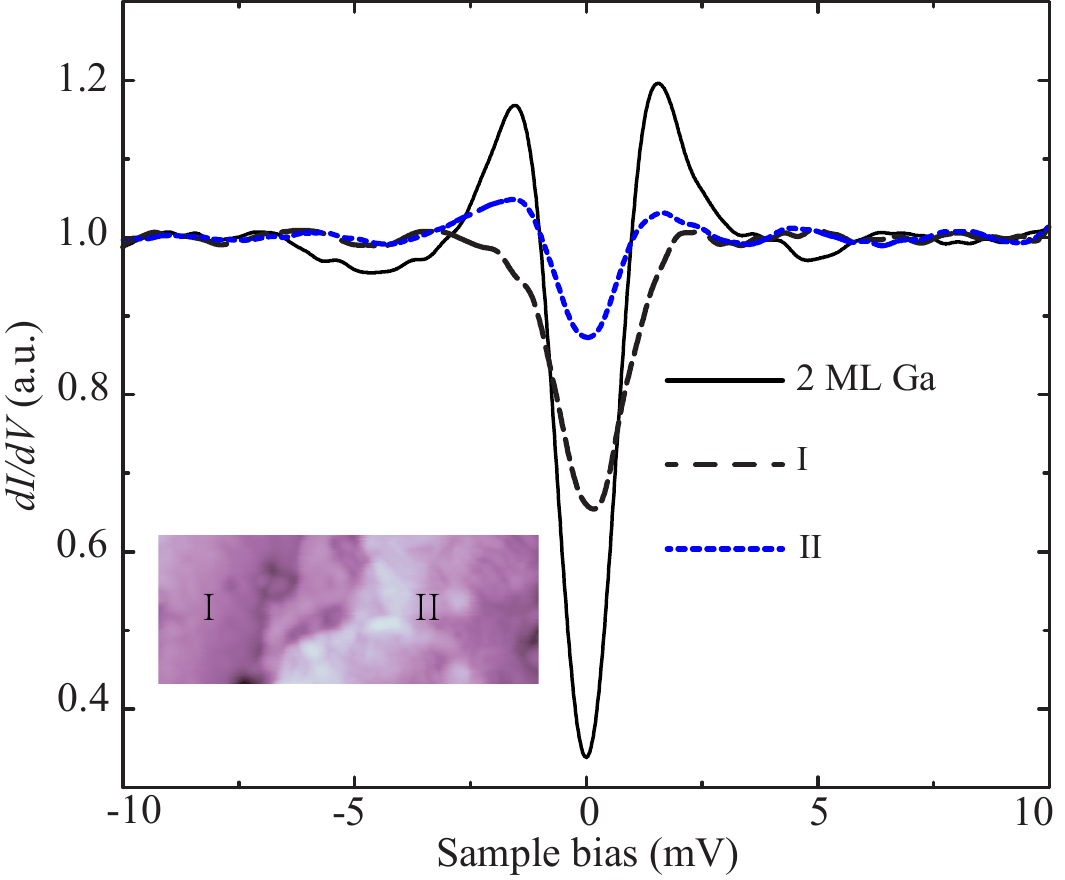}
\caption{(Color online). Normalized \textit{dI/dV }spectra before and after Ag deposition. Inset shows 2 ML Ga (I) partially covered by 1 ML Ag (II) (1.8 V, 0.05 nA, 25 $\times$ 10 nm$^2$). Three spectra are acquired at pristine Ga films (solid black curve), regions I (black dashes) and II (blue dashes), respectively.
}
\end{figure}

We now comment on the role of Ag capping layer for\textit{ ex situ} transport and magnetization measurements. One may wonder whether the metallic Ag will suppress the superconductivity of 2 ML Ga films due to proximity effects between them \cite{Cooper1961, Long2004, Sternfeld2005, Bose2007}. Indeed, as demonstrated in Fig.\ 5, the proximity effect induced superconductivity in Ag layer (blue dashes) accompanied with the supressed superconductivity in Ga films are clearly identified by comparing\textit{ dI/dV} spectra before (black curve) and after (black and blue dashes) Ag deposition. Previous study has revealed that $T_{\textrm{c}}$ of a superconductor/normal metal bi-layer system decays exponentially with the thickness ratio of normal metal and superconductor \cite{Cooper1961}. In our case, 2 ML Ga films are only 0.552 nm thick, while the Ag capping layer 80 nm thick. Thus, if the capping layer Ag is in good contact with Ga films, it would mean an almost disappearance of superconductivity in Ag/Ga bilayer. The robust superconductivity observed here therefore appears quite unexpected and surprising. Two possible causes might be considered. First, the ultrathin Ga films are not freestanding, but supported by a GaN substrate, which may help maintain the superconductivity. Second, the capping Ag layer exists in the form of nanoparticles with a typical size of $\sim$ 10 nm, compariable to $\xi_{ab} \sim10$ nm deduced from $B_{c2}$ $\sim$ 3.26 T by $B_{\textrm{c2}}$=$\Phi_{0}/2\pi\xi_{ab}^2$. Therefore, not all Ga films contact directly with Ag. Instead the Ag nanoparticles will pile up together and leave many vacant spaces among them as well as between them and the underlying Ga films. As a consequence, the proximity effect develops only in a tiny minority of regions with Ag contacting the underlying 2 ML Ga films, leaving most other regions little affected. These regions may percolate through the whole films and form an infinite superconducting percolating network, which finally leads to the superconducting behaviors detected by transport measurements and BKT-like transition \cite{Wysin2005}. Further theoretical and experimental investigations in this context would be helpful to wholly pin down this issue.

Finally, we remark on the possible mechanism of high $T_{\textrm{c}}$ in 2 ML Ga/GaN(0001) hybrid heterostructure. One may expect that the dimensionality effect plays a role. However, previous studies have revealed the strongly suppressed superconductivity as a superconductor gets thinner \cite{Brun2009, Song2011}. It is therefore unlikely that the observed high $T_{\textrm{c}}$ of 5.4 K stems solely from the dimensionality effect. In analogous to recent studies \cite{Zhang2010, Wang2012, Zhang2014}, we suggest that the superconductivity observed here may originate from the interface effect between Ga and GaN. The almost same lattice constant infers the possible strong interface interactions between Ga and GaN. On the other hand, GaN has a noncentrosymmetric crystal structure and may exhibit strong polarization effect \cite{Bernardini1997}, which helps prompt superconductivity at the Ga/GaN interface. In fact, the 2D electron gas, a prerequisite for superconductivity, is indeed observed in wurtzite hetereostructures such as AlGaN/GaN \cite{Ambacher1999}, which has been explained as polarization effect induced interface charge accumulation.

In summary, 2 ML Ga films with hexagonal atomic structure has been successfully grown on GaN(0001) substrate, and demonstrated to be a 2D superconductor by both \textit{in situ} STM/STS and \textit{ex situ} electrical magneto-transport and magnetization measurements. Compared to stable $\alpha$-Ga phase, $T_{\textrm{c}}$ in 2 ML Ga/GaN(0001) is considerably enhanced. Our finding may provide a 2D system for uncovering the nature of interface superconductivity.

\begin{acknowledgments}
H. M. Zhang, Y. Sun and W. Li contributed equally to this work. This work was financially supported by National Basic Research Program of China and the National Natural Science Foundation of China. All STM images were processed by Nanotec WSxM software \cite{horcas2007wsxm}.
\end{acknowledgments}

%

\end{document}